\documentclass[twocolumn]{aastex63}
\usepackage{showyourwork}
\DeclareRobustCommand{\Eqref}[1]{Eq.~\ref{#1}}
\DeclareRobustCommand{\Figref}[1]{Fig.~\ref{#1}}

\usepackage{flushend}
\usepackage{amsmath}
\graphicspath{{./figures/}}

\begin{document}

% Title
\title{Pair-instability mass loss for top-down compact object mass calculations}

% Author list
\author[0000-0002-6718-9472]{M.~Renzo}
\affiliation{Center for Computational Astrophysics, Flatiron Institute, New York, NY 10010, USA}
\affiliation{Department of Physics, Columbia University, New York, NY 10027, USA}

\author[0000-0002-8717-6046]{D.~D.~Hendriks}
\affiliation{Department of Physics, University of Surrey, Guildford, GU2 7XH, Surrey, UK}

\author[0000-0001-5484-4987]{L.~A.~C.~van~Son}
\affiliation{Center for Astrophysics $|$ Harvard $\&$ Smithsonian,60 Garden St., Cambridge, MA 02138, USA}
\affiliation{Anton Pannekoek Institute for Astronomy, University of Amsterdam, Science Park 904, 1098XH Amsterdam, The Netherlands}
\affiliation{Max-Planck-Institut für Astrophysik, Karl-Schwarzschild-Straße 1, 85741 Garching, Germany}

\author[0000-0003-3441-7624]{R.~Farmer}
\affiliation{Max-Planck-Institut für Astrophysik, Karl-Schwarzschild-Straße 1, 85741 Garching, Germany}

\begin{abstract}
  \noindent
  Population synthesis relies on semi-analytic formulae to determine
  masses of compact objects from the (helium or carbon-oxygen) cores of
  collapsing stars. Such formulae are combined across mass
  ranges that span different explosion mechanisms, potentialy
  introducing artificial features in the compact object mass
  distribution. Such artifacts impair the interpretation of
  gravitational-wave observations. We propose a
  ``top-down'' remnant mass prescription where we remove mass from the
  star for each possible mass-loss mechanism, instead of relying on the fallback
  onto a ``proto-compact-object'' to get the final mass. For one of these
  mass-loss mechanisms, we fit
  the metallicity-dependent mass lost to pulsational-pair instability supernovae
  from numerical simulations. By imposing no mass loss in the absence of
  pulses, our approach recovers the existing compact object masses
  prescription at the low mass end and ensures continuity across the
  core-collapse/pulsational-pair-instability regime. Our remnant mass
  prescription can be extended
  to include other mass-loss mechanisms at the final collapse.\\
\end{abstract}

\section{Introduction}

Stellar and binary population synthesis calculations are necessary to
predict event rates and population statistics of astrophysical
phenomena, including those involving neutron stars (NS) and
black holes (BH). Typically, at the end of the evolution (carbon
depletion) the mass of the core is mapped to a compact object mass, using a
$M_\mathrm{comp.\ obj}\equiv M_\mathrm{comp.\ obj}(M_\mathrm{core})$
informed by core-collapse (CC) simulations (e.g., \citealt{fryer:12,
  spera:15, mandel:20, couch:20}, see also \citealt{zapartas:21,
  patton:21}) and/or (pulsational) pair instability (PPI) simulations
(e.g., \citealt{belczynski:16, woosley:17, spera:17, stevenson:19,
  marchant:19, farmer:19, breivik:20, renzo:20csm, costa:21}).

The most commonly adopted algorithms to obtain compact object masses
in the CC regime are the ``rapid'' and ``delayed''
prescriptions of \cite{fryer:12}. In both cases, the compact object
is built from the ``bottom-up'', starting from a proto-NS mass and adding the amount of
fallback expected in the (possibly failed) explosion. However the
proto-NS mass and information about the core structure relevant
to decide the fallback are usually not available in population
synthesis calculations \citep[e.g.,][]{patton:20}. Instead the total final mass of
the star is arguably easier to constrain in population synthesis models.

At the transition between CC and PPI (roughly at carbon-oxygen cores
of $\sim{}35\,M_\odot$, \citealt{woosley:17, marchant:19, farmer:19, renzo:20csm,
  costa:21}), a mismatch between commonly adopted fitting formulae
exists. These are at least partly caused by differences in the stellar structures used to
design these algorithms and impair the
interpretation of gravitational-wave data \citep[as pointed out in
Fig.~5 of][]{vanson:21}. While it is possible that the BH mass
function is discontinuous at the onset of the PPI regime (e.g.,
\citealt{renzo:20conv,costa:21}, Hendriks et al., in prep.), the
location and amplitude of a putative discontinuity should not be
governed by a mismatch between the fitting formulae from different stellar
evolution models.

% \vspace*{-20pt}
\section{Top-down compact object masses}

In contrast with the  ``bottom up'' approach of
\cite{fryer:12}, we propose a ``top-down'' way to build the compact object masses
$M_\mathrm{comp.\ obj}$. Starting from the total stellar mass we
remove the amount of mass lost due to all of the processes
associated with the (possibly failed) explosion:

\begin{widetext}
  \begin{equation}
    \label{eq:mass}
      M_\mathrm{comp.\ obj} =
      M_\mathrm{pre-CC} - \left(\Delta M_\mathrm{SN} + \Delta M_{\nu, \mathrm{core}} + \Delta M_\mathrm{env} + \Delta M_\mathrm{PPI} + \cdots \right)
  \end{equation}
\end{widetext}

where all masses are in $M_\odot$ units, $M_\mathrm{pre-CC}$ is the
total mass at the onset of CC, and each term in the parenthesis
corresponds to a potential mass-loss mechanism: $\Delta M_\mathrm{SN}$
for the CC ejecta, $\Delta M_{\nu, \mathrm{core}}$ the change in
gravitational mass of the core due to the neutrino losses,
$\Delta M_\mathrm{env}$ the loss of the envelope due to the change in
gravitational mass corresponding to $\Delta M_{\nu, \mathrm{core}}$
that can occur even in red supergiant ``failed'' core-collapse
\citep{nadezhin:80, lovegrove:13, piro:13, fernandez:18, ivanov:21},
and $\Delta M_\mathrm{PPI}$ the pulsational mass loss due to
pair-instability. Each term may further be a function of the progenitor stellar or binary
properties, and may be theoretically or observationally informed
(e.g., $\Delta M_\mathrm{SN}$ could be derived from the light curves of a
large sample of observed SNe). \Eqref{eq:mass} can be further extended by
adding additional mass-loss mechanisms in the parenthesis (e.g., disk
winds).

In the CC regime the previous approach from \cite{fryer:12} can be recovered by setting
$\Delta M_\mathrm{SN} + \Delta M_{\nu, \mathrm{core}} = M_\mathrm{pre-CC} - M_\mathrm{comp.\ obj}^\mathrm{Fryer+12}$,
where the last term is the compact object mass as predicted by
\cite{fryer:12} and ignoring the other mass loss terms,
such as $\Delta M_\mathrm{PPI}$ and $\Delta M_\mathrm{env}$.

\section{New fit for PPI ejecta}

\begin{figure}[bp]
    \begin{centering}
      \includegraphics[width=0.75\linewidth]{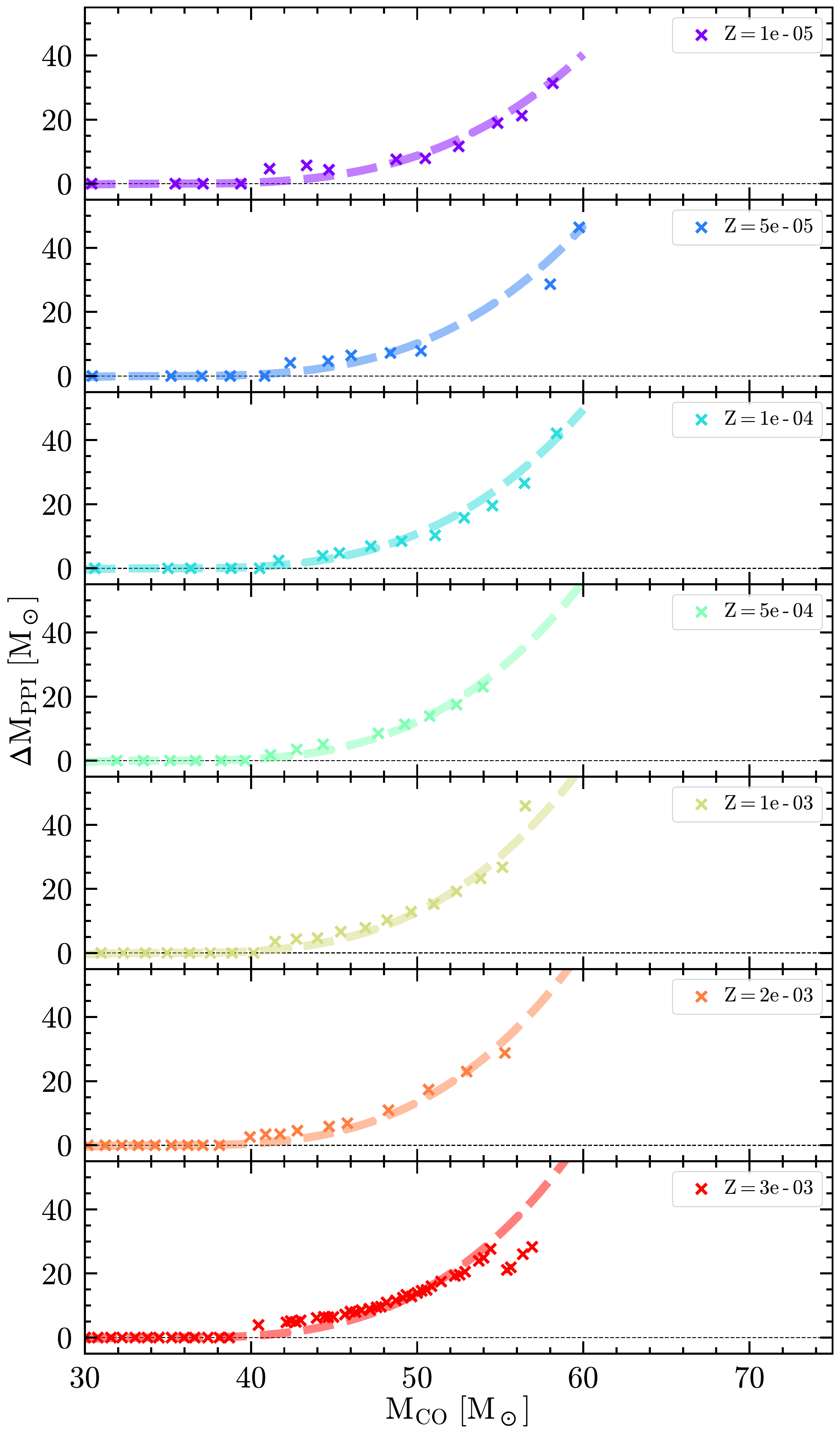}
      \caption{Each panel shows our fitting formula \Eqref{eq:fit} for
        the amount of PPI induced mass-loss as a function of carbon-oxygen
        core mass at each metallicity $Z$ as computed in
        \cite{farmer:19}. The crosses show the values from Tab.~1
        \cite{farmer:19}.}
        % This label tells showyourwork that the script `figures/fit_DM_PPI.py'
        % generates the PDF included above
        \label{fig:fit_DM_PPI}
    \end{centering}
\end{figure}

The top-down approach of \Eqref{eq:mass}, imposes
$\Delta M_\mathrm{PPI}=0$ at the edge of the PPI regime,
which produces a smooth BH mass
distribution. We fit the metallicity-dependent naked helium core PPI
simulations of \cite{farmer:19} to obtain
$\Delta M_\mathrm{PPI} \equiv \Delta M_\mathrm{PPI}(M_\mathrm{CO},Z)$ as
a function of the carbon-oxygen core mass (\Figref{fig:fit_DM_PPI}). While the fit of
\cite{farmer:19} provides the remaining mass after PPI, this is only
an estimate of the BH mass because of the the other mass loss processes that
might occurr \citep[e.g.,][]{renzo:20csm, powell:21, rahman:22}. Our approach
fits the mass removed by PPI only, which is what is directly computed
in \cite{farmer:19}.

The dashed curves in each panel of \Figref{fig:fit_DM_PPI} show the
fit \Eqref{eq:fit} for each metallicity computed in
\cite{farmer:19}. We neglect the (weak) metallicity dependence of the
minimum core mass for PPI, and we fit the publicly available data for initial He core
masses between $38-60\,M_\odot$.  We emphasize that \cite{farmer:19}
only simulated helium cores. In the case of a star with a H-rich envelope which is still
present at the onset of the pulsations and if it is both extended and red, it can be
easily removed by the first pulse
\citep[][]{woosley:17,renzo:20csm}. Thus the H-rich mass of red supergiants should be added to the
$\Delta M_\mathrm{PPI}$ provided here. It is unclear what occurs in
cases when the envelope is compact and blue \citep[e.g.,][]{dicarlo:19, renzo:20merger, costa:21}.

\begin{widetext}
\begin{equation}
\label{eq:fit}
\Delta M_\mathrm{PPI} = (0.0006\log_{10}(Z)+0.0054)\times (M_\mathrm{CO}-34.8)^3-0.0013\times (M_\mathrm{CO}-34.8)^2
\end{equation}
\end{widetext}
 %% generated by src/figures/fit_DM_PPI.py

The amount of mass lost in PPI is sensitive to convection inside the star
\citep{renzo:20conv} and the assumed nuclear reaction rates \citep{farmer:19,
  farmer:20, costa:21, woosley:21, mehta:21}, which can introduce
  uncertainties up to $\sim{}20\%$ on the maximum BH mass. The accuracy
of our fit is comparable to these uncertainties.

%\vspace*{-10pt}
\section*{Acknowledgements}
MR is grateful to R.~Luger for help with showyourwork
\citep{luger:21}. The code associated to this paper is publicly
available at
\url{https://github.com/mathren/top_down_compact_obj_mass} and the
input data are loaded from \url{https://zenodo.org/record/3346593}.  LvS
acknowledges partial financial support from the National Science
Foundation under Grant No. (NSF grant number 2009131), the Netherlands
Organisation for Scientific Research (NWO) as part of the Vidi
research program BinWaves with project number 639.042.728 and the
European Union’s Horizon 2020 research and innovation program from the
European Research Council (ERC, Grant agreement No. 715063).

\newpage
\bibliography{./top_down_comp_obj_mass.bib}
\end{document}